\documentstyle[psfig]{mn}

\newif\ifAMStwofonts

\newcommand{\1}{\begin{equation}}
\newcommand{\2}{\end{equation}}
\newcommand{\3}{\begin{eqnarray}}
\newcommand{\4}{\end{eqnarray}}

\newcommand{\rs}{ redshift }

\ifoldfss

  \newcommand{\bld}[1] {{\bf #1}}
  \ifCUPmtlplainloaded \else
    \NewTextAlphabet{textbfit} {cmbxti10} {}
    \NewTextAlphabet{textbfss} {cmssbx10} {}
    \NewMathAlphabet{mathbfit} {cmbxti10} {} 
    \NewMathAlphabet{mathbfss} {cmssbx10} {} 
  \fi
  \ifAMStwofonts
    \ifCUPmtlplainloaded \else
      \NewSymbolFont{upmath} {eurm10}
      \NewSymbolFont{AMSa} {msam10}
      \NewMathSymbol{\upi}     {0}{upmath}{19}
      \NewMathSymbol{\umu}     {0}{upmath}{16}
      \NewMathSymbol{\upartial}{0}{upmath}{40}
      \NewMathSymbol{\leqslant}{3}{AMSa}{36}
      \NewMathSymbol{\geqslant}{3}{AMSa}{3E}

    \fi
  \fi
\fi 

\ifnfssone
  \newmathalphabet{\mathit}
  \addtoversion{normal}{\mathit}{cmr}{m}{it}
  \addtoversion{bold}{\mathit}{cmr}{bx}{it}

  \newcommand{\bld}[1] {\mathbf{#1}}
  \newmathalphabet{\mathbfit} 
  \addtoversion{normal}{\mathbfit}{cmr}{bx}{it}
  \addtoversion{bold}{\mathbfit}{cmr}{bx}{it}
  \newmathalphabet{\mathbfss} 
  \addtoversion{normal}{\mathbfss}{cmss}{bx}{n}
  \addtoversion{bold}{\mathbfss}{cmss}{bx}{n}
  \ifAMStwofonts
    \ifCUPmtlplainloaded \else
      \UseAMStwoboldmath
      \makeatletter
      \new@mathgroup\upmath@group
      \define@mathgroup\mv@normal\upmath@group{eur}{m}{n}
      \define@mathgroup\mv@bold\upmath@group{eur}{b}{n}
      \edef\UPM{\hexnumber\upmath@group}
      \new@mathgroup\amsa@group
      \define@mathgroup\mv@normal\amsa@group{msa}{m}{n}
      \define@mathgroup\mv@bold\amsa@group{msa}{m}{n}
      \edef\AMSa{\hexnumber\amsa@group}
      \makeatother
      \mathchardef\upi="0\UPM19
      \mathchardef\umu="0\UPM16
      \mathchardef\upartial="0\UPM40
      \mathchardef\leqslant="3\AMSa36
      \mathchardef\geqslant="3\AMSa3E
    \fi
  \fi
\fi 

\ifnfsstwo

  \newcommand{\bld}[1] {\mathbf{#1}}
  \DeclareMathAlphabet{\mathbfit}{OT1}{cmr}{bx}{it}
  \SetMathAlphabet\mathbfit{bold}{OT1}{cmr}{bx}{it}
  \DeclareMathAlphabet{\mathbfss}{OT1}{cmss}{bx}{n}
  \SetMathAlphabet\mathbfss{bold}{OT1}{cmss}{bx}{n}
  \ifAMStwofonts
    \ifCUPmtlplainloaded \else
      \DeclareSymbolFont{UPM}{U}{eur}{m}{n}
      \SetSymbolFont{UPM}{bold}{U}{eur}{b}{n}
      \DeclareSymbolFont{AMSa}{U}{msa}{m}{n}
      \DeclareMathSymbol{\upi}{0}{UPM}{"19}
      \DeclareMathSymbol{\umu}{0}{UPM}{"16}
      \DeclareMathSymbol{\upartial}{0}{UPM}{"40}
      \DeclareMathSymbol{\leqslant}{3}{AMSa}{"36}
      \DeclareMathSymbol{\geqslant}{3}{AMSa}{"3E}
    \fi
  \fi
\fi 

\ifCUPmtlplainloaded \else
  \ifAMStwofonts \else 
    \def\upi{\pi}
    \def\umu{\mu}
    \def\upartial{\partial}
  \fi
\fi

\title{PINOCCHIO and the hierarchical  build-up of dark matter haloes}
\author[Taffoni, Monaco \& Theuns]
       {Giuliano Taffoni$^1$, Pierluigi Monaco$^2$ \& Tom Theuns$^3$ \\
        $^1$SISSA, Via Beirut 2-4, 34060 Trieste, Italy\\
        $^2$Dipartimento di Astronomia, via Tiepolo 11, 34131 Trieste, Italy\\
        $^3$Institute of Astronomy, Madingley Road, Cambridge CB3 0HA, UK}
\date{Accepted 0000 December 1.
      Received 0000 December 1;
      in original form 0000 October 1}

\pagerange{\pageref{firstpage}--\pageref{lastpage}}
\pubyear{0000}

\begin{document}
\maketitle

\label{firstpage}

\begin{abstract}
  We study the ability of PINOCCHIO (PINpointing Orbit-Crossing
  Collapsed HIerarchical Objects) to predict the merging histories of
  dark matter (DM) haloes, comparing the PINOCCHIO predictions with the
  results of two large N-body simulations run from the same set of
  initial conditions.  We focus our attention on quantities most
  relevant to galaxy formation and large-scale structure studies.
  PINOCCHIO is able to predict the statistics of merger trees with a
  typical accuracy of 20 per cent.  Its validity extends to
  higher-order moments of the distribution of progenitors.  The
  agreement is valid also at the object-by-object level, with 70-90
  per cent of the progenitors cleanly recognised when the parent halo
  is cleanly recognised itself.  Predictions are presented also for
  quantities that are usually not reproduced by semi-analytic codes,
  such as the two-point correlation function of the progenitors of
  massive haloes and the distribution of initial orbital parameters of
  merging haloes.  For the accuracy of the prediction and for the
  facility with which merger histories are produced, PINOCCHIO
  provides a means to generate catalogues of DM haloes which is
  extremely competitive to large-scale N-body simulations, making it a
  suitable tool for galaxy formation and large-scale structure
  studies.
\end{abstract}


\section{Introduction}

In the {\em Hierarchical Clustering Scenario}, structure in the
Universe forms from the aggregation and merging of smaller subunits.
This theoretical picture is now substantiated, at least on a
qualitative level, by a wealth of observations of the high-redshift
($z\sim3-5$) Universe.  In the most commonly discussed scenario, the
\lq Cold Dark Matter\rq\ (CDM) one, hierarchical clustering is driven
by the gravitational collapse of DM fluctuations, while the visible
astrophysical objects are generated from baryons falling into the DM
haloes (see, e.g., White \& Rees 1978).  Thus the process of formation
and evolution of DM haloes is of fundamental importance for
understanding the properties of galaxies or galaxy clusters.

The formation of DM haloes involves highly non-linear dynamical
processes which can not be followed analytically. To face this
problem it is necessary to resort to numerical N-body simulations.  
Besides this time-consuming method, one can use also analytical 
approximations  that are able to
predict with fair accuracy some relevant quantities related to the
assembly of DM haloes.  Moreover, the analytic methods help to shead
light on the complex gravitational problem of hierarchical
clustering.  The pioneers of the analytical approach were Press \&
Schechter (1974; hereafter PS) who derived an expression for the mass
function of DM haloes.  This was found to give a fair approximation of
the N-body results (Efstathiou et al. 1988; see for a review Monaco 1998). 
The PS approach was extended by Bond et al. (1991; see also
Peacock \& Heavens 1990; Bower 1991; Lacey \& Cole 1993) 
(Extended PS formalism, hereafter EPS), who fixed a
normalization problem of the original PS work.  
The EPS model can be used to predict also some
properties of DM haloes, such as their formation time,  survival time 
and merger rate.  These predictions were tested
against numerical simulations, again with success, by Lacey \& Cole
(1994).  The EPS formalism has recently become a standard tool to
construct synthetic catalogues of DM haloes for galaxy formation
programs (see, e.g., Kauffmann, White \& Guiderdoni  1993;
 Somerville \& Primack 1999; Cole et al. 2000).

However, recent work with larger N-body simulations has revealed
significant discrepancies between PS and EPS predictions and numerical
results.  The PS mass function has been shown to underpredict the
number of massive haloes and over-predict the number of low-mass
ones (Gelb \& Bertschinger 1994, Governato et al. 1999; Jenkins et al. 2001; 
Bode et al. 2001).
Similar discrepancies were observed in the reconstruction of the
conditional mass function, i.e. the number density of haloes
bound to flow into a parent halo of given mass at a subsequent time
\footnote{In the following, the \lq final\rq\ haloes at
$z=0$ (or occasionally at higher redshift) 
will be called {\it parent}, while the higher-redshift haloes 
that flow into the parent will be called {\it progenitors}.}  
(Somerville \& Kolatt 1999; Sheth \& Lemson 1999b).  The EPS formalism
is also affected by limitations, in that it does not give full
information on the spatial distribution of haloes (Catelan et al. 1998;
Jing 1998, 1999; Porciani, Catelan \& Lacey 1999), and by
inconsistencies in the use of smoothing filters and in the
construction of merger trees (Somerville \& Kolatt 1999; Sheth \&
Lemson 1999b; Cole et al. 2001).  Attempts to improve this formalism,
or to develop alternative ones, were reviewed by Monaco (1998).  A
more recent and successful extension is due to Sheth \& Tormen (1999)
and Sheth, Mo \& Tormen (2001); their model improves significantly the
fit of the mass function and the extension of dynamics to ellipsoidal
collapse (EPS is based on linear theory), but does not remove the
inconsistencies of the EPS approach and does not provide spatial
information of haloes either.  This method has also been applied to
build random realizations of the merging histories of DM haloes (Sheth
\& Tormen 2001), but it does not provide a significant improvement
respect to the standard merger trees.

Recently, we have presented a new algorithm, called PINOCCHIO
(PINpointing Orbit-Crossing Collapsed HIerarchical Objects), to
generate synthetic catalogues of DM haloes with known mass, position,
peculiar velocity, merger history and angular momentum (Monaco et al.
2001, hereafter paper I; Monaco, Theuns \& Taffoni 2001, hereafter
paper II).  In contrast to EPS, PINOCCHIO is able both to reproduce
statistical quantities, such as the mass or two-point correlation
function of haloes, and to reproduce haloes on an point-by-point basis.

In this paper, we investigate in detail how well PINOCCHIO is able to
recover the merger histories (or {\em merger trees}) of DM haloes.  We
compare the PINOCCHIO code with numerical N-body simulations and with
the analytical extimates of the EPS theory.  We examine the ability of
PINOCCHIO to recostruct the main statistical properties of the
merger trees, extending the analysis to predict the correlation
function and the initial orbital parameters of merging haloes.

Section~2 gives a brief description of the PINOCCHIO code with special
attention to the extraction of the merger trees.  In Section~3 we
compare the statistical properties of the distribution of DM haloes at
different redshifts given by PINOCCHIO with the results of numerical
N-body simulations.  Section~4 is dedicated to the study of the
spatial distribution of the haloes that will form cluster sized objects
at the present time.  Section~5 shows the ability of PINOCCHIO to
predict the impact parameters of merging haloes.  The conclusions are
reported in Section~6.
\begin{figure*}
\label{fig:NmLCDM}
\centerline{
\psfig{file=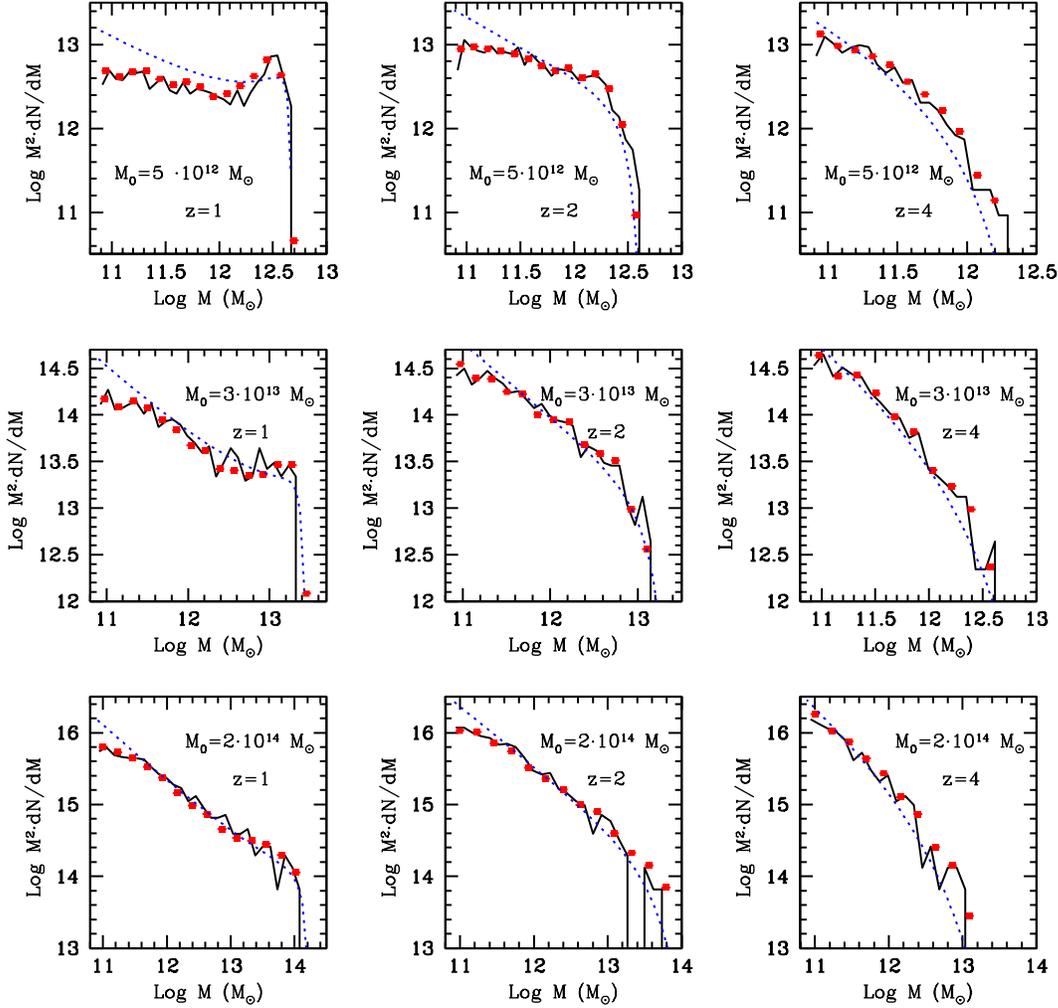,width=15cm,height=15cm}}
\caption{Conditional mass functions 
  in the $\Lambda$CDM case for parent haloes identified at $z=0$.  The
  mass threshold is fixed at $M_{\rm th}=7.6 \times 10^{10}$ M$_\odot
  $ (10 particles), the \rs increases from left to right and covers
  the values: $z=1, \; 2,\; 4$.  The mass of the parent halo increases
  from top to bottom, the adopted values are: $M_0 = 5. \times 10^{12}
  M_\odot$, $3. \times 10^{13} M_\odot$ and $2.0 \times 10^{14}
  M_\odot$.  The points represent the simulation data while the solid
  lines are the prediction of PINOCCHIO; the dashed lines are the
  analytical predictions of the EPS formalism.}
\end{figure*}
%

\section{Merger trees from PINOCCHIO}

The PINOCCHIO code was presented in paper I and described in full
detail in paper II.  Here we give only a brief description of the
code, necessary to discuss the procedure used to extract the merger
trees.

For a given cosmological background model and a power spectrum of
fluctuations, a Gaussian linear density contrast field $\delta_{\rm l}$
(i.e. linearly extrapolated to $z=0$) is generated on a cubic grid, in
a way much similar to what is usually done to generate initial
condition for N-body simulations.  The linear density contrast
$\delta_{\rm l}$ is smoothed repeatedly with Gaussian filters of FWHM $R$,
where $R$ takes values that are equally spaced ($\sim$20 smoothing
radii usually give an adequate sampling).  For each point $\bld{q}$ of
the Lagrangian (initial) coordinate and for each smoothing radius $R$,
the collapse time (i.e. the time at which the particle is predicted to
enter a high-density, multi-stream region) is computed using
Lagrangian Perturbation Theory (hereafter LPT; see e.g. Bouchet 1996,
Buchert 1996, Catelan 1995) and its ellipsoidal truncation (Monaco
1997).  Technically, the collapse time is defined as the instant of
Orbit Crossing (OC); this definition is discussed at length in paper
II (see also Monaco 1995, 1997a).  For each particle only the earliest
collapse time is recorded which amounts to recording the field

\begin{equation}
F_{\rm max}(\bf{q}) \equiv \max_{R} \left( \frac{1}{b(t_{\rm c})} \right).
\label{eq:fmax}
\end{equation}

\noindent
Here $b(t)$ is the linear growing mode (see Padmanabhan 1993; Monaco
1998) and $t_{\rm c}$ is the OC-collapse time.  Notice that in an
Einstein-de Sitter Universe $F_{\rm max} = (1+z_{\rm max})$, where
$z_{\rm max}$ is the largest collapse redshift at which the particle 
collapses.\footnote
{Taking the largest redshift (or $F$-value) of collapse is analogous
to considering the largest collapse radius, as done in the EPS
formalism.}
%
\begin{figure*}
\label{fig:NmSCDM}
\centerline{
\psfig{file=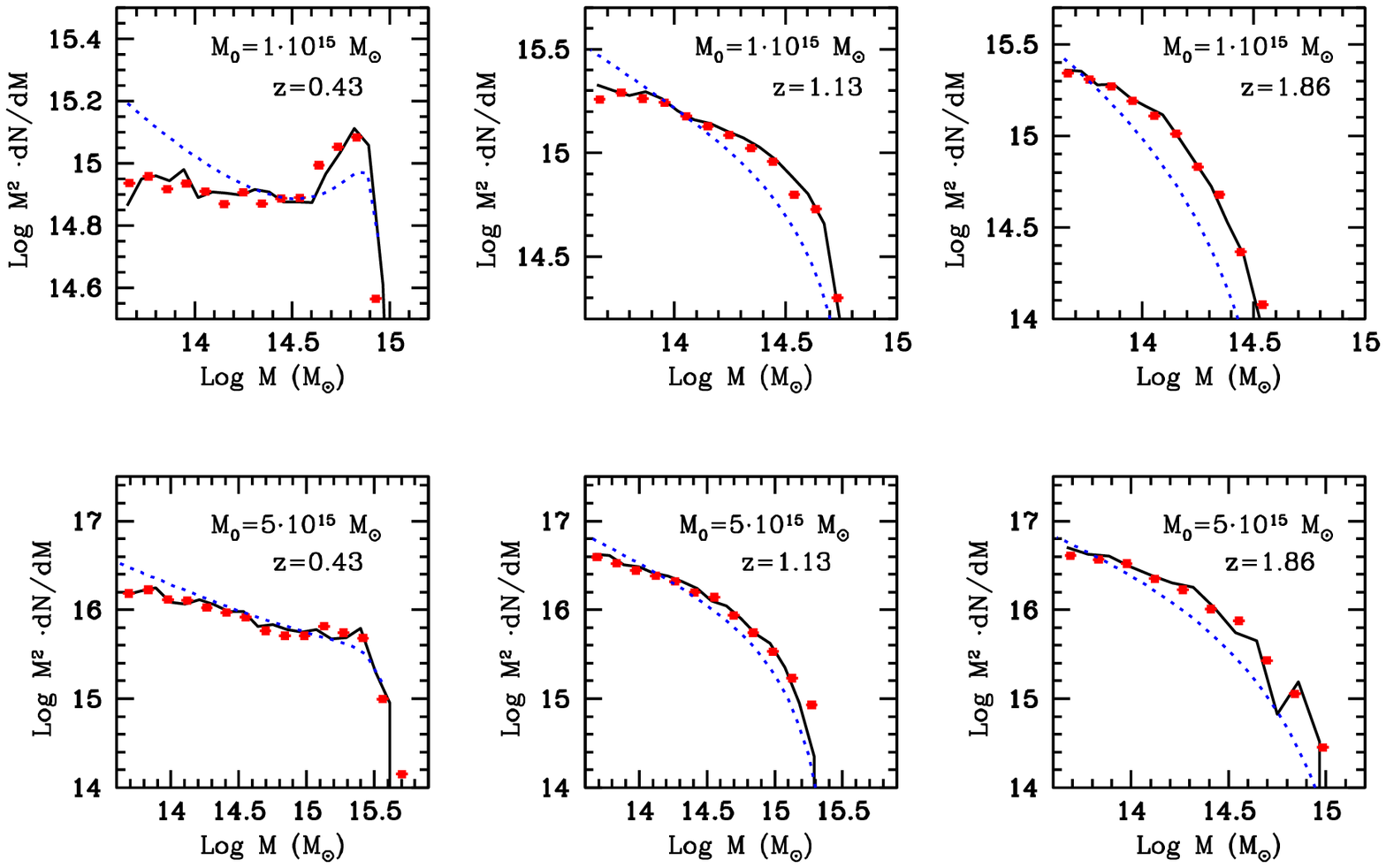,width=15cm,height=11cm}}
\caption{Same as  in Fig.~1 but for  the  SCDM case. 
  The mass threshold is $M_{\rm th}=1.49 \times 10^{13} M_\odot $ (10
  particles).  }
\end{figure*}
%

Besides $F_{\rm max}$, we record also the smoothing radius $R_{\rm
  max}$ for which the maximum in equation~(\ref{eq:fmax}) is reached,
and the velocity of the particle at collapse time as given by the
Zel'dovich approximation (1970) (which is the first-order term of
LPT),

\begin{equation}
{\bf v}_{\rm max} = - b(t) \nabla \phi({\bf q};R_{\rm max})
\label{eq:zelvel}
\end{equation}

\noindent
(in units of comoving displacement), where $\phi({\bf q};R_{\rm max})$
is the rescaled peculiar gravitational potential (smoothed at $R_{\rm
  max}$), which obeys the Poisson equation $\nabla^2 \phi ({\bf q}) =
\delta_{\rm l}$.  All differentiations and convolutions are performed using
fast Fourier transforms.

The collapsed medium is then ``fragmented'' into isolated objects
using an algorithm designed to mimic the accretion and merger events
of hierarchical collapse.  Collapsed particles may belong to relaxed
haloes or to lower-density filaments.  At the instant a particle is
deemed to collapse, we decide which halo, if any, it accreted onto.
The candidates haloes are those that already contain one Lagrangian
neighbour of the particle (on the initial grid ${\bf q}$ of Lagrangian
positions, the six particles nearest to a given one are its
``Lagrangian neighbours'').  The particle will accrete onto the halo
if its distance (in the Eulerian space at the collapse time) from the
centre of mass of the candidate halo is smaller than a given fraction
of the halo size ($R_N=N^{1/3}$ in grid unit, where $N$ is the number
of particles in the halo), otherwise they are catalogued as filaments.
If a particle has more than one candidate halo, we check whether these
haloes should merge. The merging condition is very similar to the
accretion one: two groups merge if their distance in the Lagrangian
space is smaller than a fraction of the size of the largest halo.  If
a particle is a local maximum of $z_{\rm max}$ it is considered as the
seed of a new halo.  Finally, filament particles are accreted onto a
halo when they neighbour in the Lagrangian space an accreting
particle; this is done to mimic accretion of filaments onto haloes.

The fragmentation algorithm requires the introduction of free
parameters, which are analogous to those required by any clump-finding
algorithm applied to N-body simulations.  These parameters specify the
level of overdensity at which the halo is defined, or are introduced
to fix resolution effects.  They are discussed in paper II (to which
we refer for all details) and chosen by requiring the fit of the mass
function of haloes selected with the friends-of-friends (FOF) algorithm
with linking length 0.2 times the mean inter-particle distance at
$z=0$.  Here we use for the parameters the values found in paper II.

To give a taste of the speed of PINOCCHIO with respect to simulations,
a 256$^3$ particles realization needs about 6 hours on a Pentium III
450MHz computer, with a RAM requirement of about 512 Mb, the
correspondig simulation required a week on 10 processor parallel
computer.  Statistical quantities like the mass function or the
two-point correlation function are reproduced with a typical error of
$\sim$10 per cent or smaller.  At the object-by-object level, the
accuracy of PINOCCHIO depends on the degree of non-linearity that is
reached at the grid level, and degrades in time.  Typically, 70-99 per
cent of the objects are reproduced with an error on the mass of 30-40
per cent, an error on the position of $\sim$ 0.5-2 grid points and a
1D error on the velocity of $\sim 150$ km/s at $z=0$.

It is noteworthy that merging events in PINOCCHIO are not restricted
to be binary: in principle up to six objects can merge together at the
same time, even if, as expected, the number of mergers that involves
more than three haloes is a very small fraction of the total.
%
\begin{table*}
\centering
\begin{minipage}{140mm}
\caption{Parameters of the considered numerical simulations}
\label{sim}
\begin{tabular}{lcccclrcccl}
Cosmology & $\Omega_0$ & $\Omega_\Lambda$ & $h_0$ &
$\sigma_8$ & n & $L_{\rm box}$ & $N_{\rm p}$ & $M_{\rm part}$ 
& output redshifts \\
\hline \\
$\Lambda$CDM & 0.3 & 0.7 & 0.65 & 0.9 & 1. & $100$Mpc/h    
& $256^3$ & $7.64 \times 10^{10} \; M_\odot$ & 0, 0.25, 0.5, 0.75, 1, 2, 3, 4, 5\\ 
SCDM & 1. & 0. & 0.5 & 1. & 1. & $500$Mpc/h   
& $360^3$  & $1.49 \times 10^{12} \; M_\odot$ & 0, 0.43, 1.13, 1.86\\ 
\end{tabular}
\end{minipage}
\end{table*}
%

The merger histories of haloes are directly evaluated by PINOCCHIO.
At each merger the largest halo retains its identification number (ID)
which will become the ID of the merger, while the other haloes are
labelled as expired. The mass of each halo involved in the merging
event is recorded together with the redshift at which the merger takes
place.  For each expired (progenitor) halo we keep track at all times
of the (parent) halo they are presently incorporated within.  Even
though accretion is rigorously defined as the entrance of a single
particle into the object, the merger of a halo with another one with
less than 10 particles is always considered as an accretion event.

The merger trees extracted from PINOCCHIO provide a more complete
description of the merging histories of haloes then the EPS one.  They
not only follow the time evolution of the mass and number distribution
of the progenitors, but also their distribution in space, their
velocities and angular momenta.
\section{Statistics of the progenitors}
\subsection{The simulations}
In order to test the ability of PINOCCHIO in predicting the statistics
of the merger trees, we compared the results of two N-body simulations
with those of PINOCCHIO applied to the same initial density field.
The simulations were already presented in paper I and II, to which we
refer for all the details.  They are a standard CDM model (SCDM), run
with the PKDGRAV code on a large box of $500\ h^{-1}$ Mpc\footnote
{The Hubble constant is assumed to be $H_0=100\ h$ km s$^{-1}$
  Mpc$^{-1}$.}
with 360$^3$ particles (Governato et al. 1999), and a $\Lambda$CDM
model, run with the Hydra code (Couchman, Thomas \& Pearce 1993) on a
smaller box of $100\ h^{-1}$ Mpc with 256$^3$ particles.  The reason
why we use two different simulations is to check the method for
different cosmologies, boxes, resolutions and codes.  For the present
purpose, the $\Lambda$CDM simulation is more suitable as the higher
mass resolution allows one to reconstruct the merger tree to higher
redshifts and lower masses, but the SCDM allows one to test the merger
trees for the more massive haloes.

The haloes are identified using a standard FOF algorithm with linking
length $0.2$ times the inter-particle distance.  Note that, following
the suggestion by Jenkins et al. (2001), we do not change linking
length with the cosmology.  In this paper, we adopt 10 particles as
the minimum mass of the haloes when we analyse the conditional mass
function.  This is to test the effect of the degrading of the
agreement at small masses. In general, at least 30 particles are
necessary to identify reliably a halo both in the simulations and in
PINOCCHIO, so we consider a threshold mass of 30 particles for the
other statistical analysis.

The merger trees for the FOF haloes at final time $z_{\rm 0}$ are
constructed as follows.  Progenitors are defined as those haloes that
at the higher redshift $z$ contain some of the particles of the parent
halo at $z_0$.  As noted by some authors (see e.g. Somerville et al.
2000), some particles that are located in a progenitor are not
included later into the parent. This reflects the actual dynamics of
the haloes that suffer stripping and evaporation events, and make the
progenitor identification process more ambiguous.  We then adopt two
simple rules:
\begin{enumerate}
\item if a parent halo contains less than 90 per cent of the mass of
  all its progenitors at redshift $z$, then it is excluded from the
  analysis (this happens in a few percent of cases);
\item we assign to the progenitor the mass of all its particles that
  will flow in to the parent at $z_0$.
\end{enumerate}
In this way we force mass conservation in the merger tree and reject
some extreme cases when the progenitor is strongly affected by these
\lq evaporation\rq\ effects.

Due to the limited number of available outputs, the merger trees
obtained from our simulations are very coarse-grained in time.
\begin{figure}
\centerline{
\psfig{file=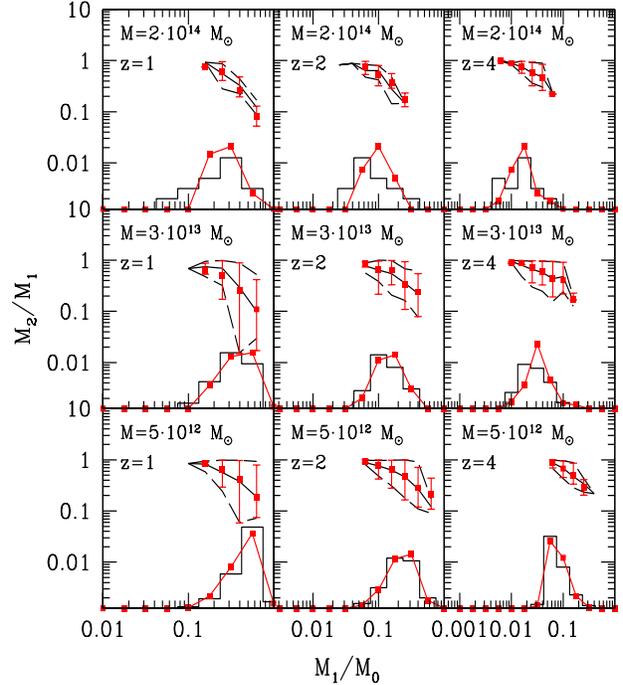,width=8.5cm,height=9.5cm}}
\caption{ The distribution of the mass of the largest
  progenitor $M_1$ for the $\Lambda$CDM case with mass threshold
  $M_{\rm th}=2.3 \times 10^{11} M_\odot $ (30 particles). The histograms
  are the PINOCCHIO predictions and the points connected with solid
  lines are the simulations'.
  The quantity plotted on the upper part of each box is the mean of
  the distribution of the mass ratio of the second largest progenitor
  $M_2$ to the first largest progenitor $M_1$ versus the mass ratio of
  the largest progenitor to the parent halo.  The solid line is the
  PINOCCHIO result and the dashed lines show its $1\sigma$ variance.
  The points with error bar are the simulation data.  }
\label{fig:m1overm0L}
\end{figure}
\begin{figure}
\centerline{
\psfig{file=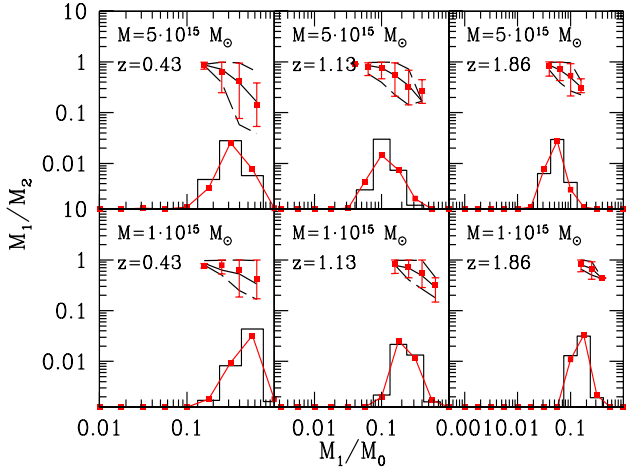,width=8.7cm,height=6.5cm}}
\caption{ As Fig.~\ref{fig:m1overm0L} but for the SCDM case. 
  The mass threshold is $M_{\rm th}=1.3 \times 10^{14} M_\odot $ 
  (30 particles).
}
\label{fig:m1overm0S}
\end{figure}
This highlights one of the advantages of using a code like PINOCCHIO
to produce the merger trees. In fact, in PINOCCHIO we follow the
merging of haloes in real time, and then we can link each progenitor to
its parent after each merging event, while in the simulations (where
haloes are identified {\em after} the run) it is necessary to analyse
and cross-correlate a large number of outputs to follow the merger
histories.  In other words, the generation of the merger trees is by
far less expensive (in term of CPU time, disk space and human labour)
in PINOCCHIO than in a simulation; in fact PINOCCHIO automatically
compute the merging history of  haloes and it does not need any
further analysis.
\subsection{Progenitor mass function}
The progenitor (conditional) mass function $dN(M,z|M_0,z_0)/dM$ is the
number density of progenitors of mass $M$ at \rs $z$ that merge to
form the parent $M_0$ at \rs $z_0$.  An estimate of this quantity
based on the PS formalism was found by Bower (1991), while Bond et al.
(1991) gave the basis for computing it in the EPS formalism (as done
by Lacey \& Cole 1993).

Let be $\delta(z)$ the critical density for a spherical perturbation
to collapse at \rs $z$ and $\Lambda(M)=\sigma^2(M)$ the variance of
the initial density field when smoothed over regions that contain on
average a mass $M$.  The fraction of mass of the parent halo that was
in the progenitors of mass $M$ at early time is: 
\3 
\nonumber
f(M,\delta|M_0,\delta_0){\rm d}M=\frac 1 {\sqrt{2 \pi}}
{(\delta-\delta_0) \over [\Lambda(M)-\Lambda(M_0)]^{3/2}} \\
\times \exp \left \{ -{(\delta-\delta_0)^2 \over
    2[\Lambda(M)-\Lambda(M_0)]}\right\} {\rm d}\,\Lambda \;, \4 and
the conditional mass function is: \1 {{\rm d} \,N \over {\rm d} \,
  M}(M,z|M_0,z_0){\rm d}M = \left (\frac {M_0}{M}\right)^2
f(M,\delta|M_0,\delta_0) \;{\rm d}M\;.
\label{eq:bower} 
\2
The PINOCCHIO conditional mass function and that obtained from
the simulations are computed by averaging over a mass interval around
$\log M_0$ of  0.01 dex. 

In Fig.~\ref{fig:NmLCDM} and Fig.~\ref{fig:NmSCDM} we compare the
conditional mass functions obtained from the EPS formalism, PINOCCHIO
and the simulations for the $\Lambda$CDM and the SCDM case
respectively.  The bottom panels of Fig. \ref{fig:NmLCDM} show for the
$\Lambda$CDM case the results for a cluster-sized parent of
$M_0=2\times 10^{14}$ M$_\odot$, the case of haloes corresponding to
small groups ($M_0=3\times 10^{13}$ M$_\odot$) and galaxies
($M_0=5\times 10^{12}$ M$_\odot$) are presented in the mid and upper
panels.  On Fig.~\ref{fig:NmSCDM} we show the results for parents with
mass comparable to massive clusters ($M_0=1\times 10^{15}$ M$_\odot$
and $M_0=5\times 10^{15}$ M$_\odot$) extracted from the SCDM
simulation.  The dotted lines show the EPS analytical prediction and
the points show the expected value computed from the simulations. The
the Poissonian errors associated to the simulation data are of the
same width of the simbols used to plot the simulation data.

The conditional mass function predicted using PINOCCHIO (the solid
lines in the plots) shows a very good agreement when compared with the
simulations.  In Fig.~\ref{fig:NmLCDM} and Fig.\ref{fig:NmSCDM} we
show that the PINOCCHIO prediction fits the simulations data with
similar accuracy for all the considered parent mass and redshifts and
we identify a discrepancy between the two distribution which in
general is less than 25 per cent.  This means that PINOCCHIO
reproduces the conditional mass function with better accuracy that the
EPS prediction and almost constant in mass and redshift.

On the other hand the figures show a discrepancy already pointed out
by other authors for the mass function of haloes (Gelb \& Bertschinger
1994; Governato et al. 1999; Jenkins et al. 2001; Bode et al. 2001 ):
the EPS prediction overestimates the number of low mass progenitors
and underestimates the number of high mass progenitors.  This
discrepancy is less evident at high redshift and it ranges from 30 per
cent to a factor of 2 or more depending on the mass of the parent
halo.
\subsection{Higher-order analysis of the progenitor distribution}
We evaluate the distribution of the mass of the largest progenitor
$M_1$ (i.e. the most massive halo that flows into the parent) for each
of the parent haloes analysed before.  The histograms on
Fig.~\ref{fig:m1overm0L} and Fig.~\ref{fig:m1overm0S} show the
distribution of the mass of the larger progenitor normalized to the
parent mass, $M_1/M_0$, predicted by PINOCCHIO for the $\Lambda$CDM
and SCDM case (in the following the mass threshold is always set to 30
particles).  The symbols connected with lines denote the corresponding
simulation results. The agreement between the numerical experiment and
PINOCCHIO is very good.  Both the mean value and the width of the
distribution are reproduced with good accuracy at all redshifts.

The distribution of $M_1/M_0$ provides also a hint on the formation
time of the parent.  In fact, the standard definition of {\em
  formation time} for a halo of mass $M_0$ is the epoch at which the
size of its largest progenitor first becomes greater than $M_0/2$.  So
we assume as the average formation redshift for a parent halo of mass
$M_0$ the time at which the peak of the distribution $M_1/M_0$ is at
one half.  The good agreement of PINOCCHIO with the simulations can
thus be extended also to the halo formation times. For instance
Fig.~\ref{fig:m1overm0S} suggests that, in this SCDM cosmology, a halo
of $1 \times 10^{15}M_\odot$ forms at $z \sim 0.43$ or later.  Notice
that a more detailed analysis of formation times is hampered by the
small number of simulation outputs available.

In the upper part of the plots of Fig.~\ref{fig:m1overm0L} and
Fig.~\ref{fig:m1overm0S} the distribution of $M_2/M_1$ (the ratio of
the second largest progenitor and largest ones) given $M_1/M_0$ is
shown.  The points are the mean value of the distribution and the
error bars are the the corresponding 1$\sigma$ variance, both measured
in the simulations.  The solid lines and the dashed lines are the same
quantities predicted by PINOCCHIO.  Again the agreement is very good.

The results reported in this section and in the previous one suggest
that the merging histories of haloes produced by PINOCCHIO reproduce
with very good accuracy the statistical properties of the masses of
those extracted from numerical simulations.  We notice that the EPS
based algorithms to produce merger trees (Lacey \& Coles 1994,
Somerville \& Kolatt 1999, Sheth \& Lemson 1999, Coles et al. 2000)
are by construction forced to reproduce the EPS analytical
distributions, and they suffer of the same discrepancy noted for the
EPS analytical prediction
(Fig.~\ref{fig:NmLCDM}~and~\ref{fig:NmSCDM}).

\subsection{The progenitors in number} 
\begin{figure*}
\centerline{
\psfig{file=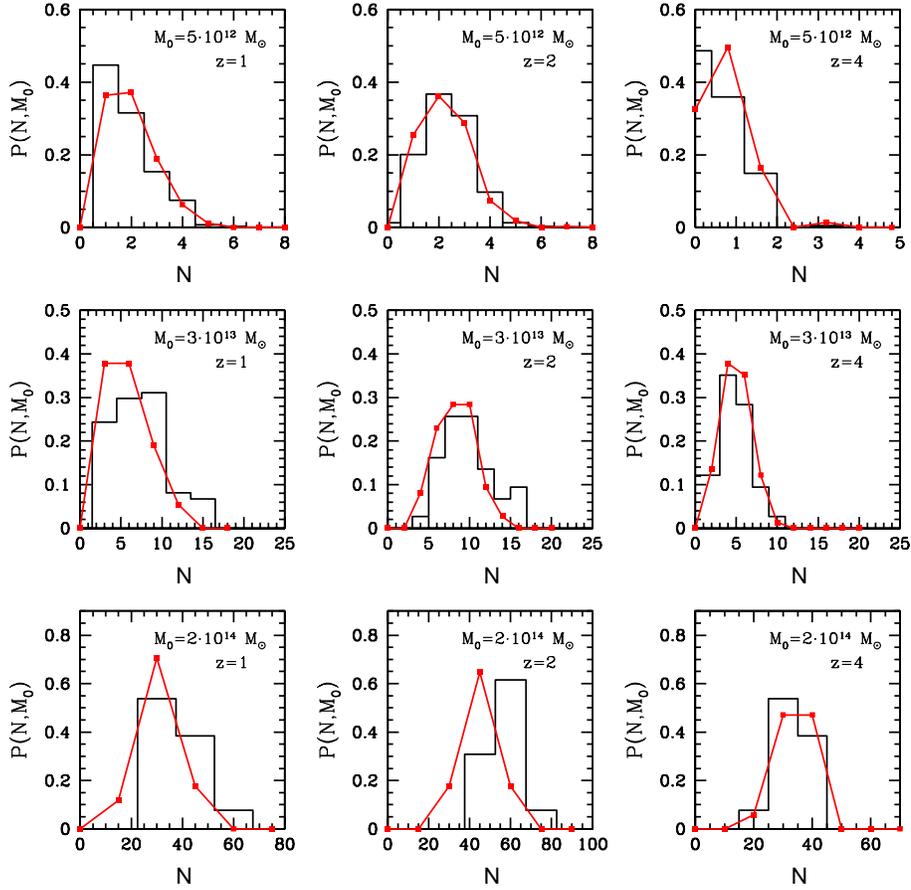,width=13cm,height=13cm}}
\caption{Probability that an halo $M_0$ at $z=0$ has
  $N$ progenitors for the $\Lambda$CDM case.  The threshold mass is
  $M_{\rm th}=2.3 \times 10^{11} M_\odot $ (30 particles). 
  The points connected with solid lines represent the simulation data
  while the histograms are the prediction of PINOCCHIO.  The vertical
  lines are the EPS analytical prediction for the mean of the
  distribution.}
\label{fig:testLn}
\end{figure*}
In this section we analyse the statistical properties of the
distribution of the number of progenitors of a halo of mass $M_0$.  

In Fig.~{\ref{fig:testLn} and Fig.~\ref{fig:testLns} we show the
  probability $P(N,M_0)$ that a halo of mass $M_0$ has $N$
  progenitors.  The average of these distribution gives (with suitable
  normalisation) the integral of the conditional mass function to the
  threshold mass, and is dominated by the more numerous small-mass
  objects.
  
  The histograms show the distribution of the number of progenitors
  evaluated from PINOCCHIO for different parent masses and redshifts.
  The filled symbols connected with lines are the distribution
  extracted from the simulation.  We notice that PINOCCHIO reproduces
  fairly well the distributions also for the more massive haloes and at
  all redshifts.
%
\begin{figure*}
  \centerline{ \psfig{file=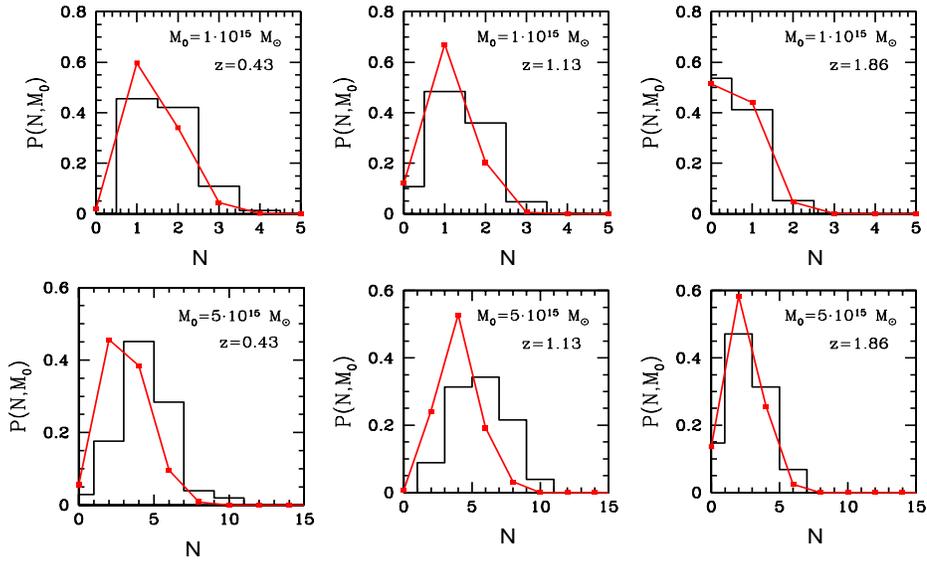,width=13cm,height=8cm}}
\caption{ Same as in Fig. 5 for the SCDM case.
  The threshold mass is $M_{\rm th}=1.3 \times 10^{14} M_\odot $ (30
  particles).
}
\label{fig:testLns}
\end{figure*}
The ability of PINOCCHIO in predicting the distribution of the number
of progenitors can be quantified by comparing the first and second
moments measured in the simulations with their values predicted by
PINOCCHIO.  In fig~\ref{fig:meanL} we show the average $\mu_1$ and the
rescaled variance $\mu_2/\mu_1$ as a function of the parent halo mass
for different redshifts.  The points connected with solid lines are
the PINOCCHIO prediction and the open symbols are the values measured form
the simulations.  The dashed lines are the EPS analytical prediction
for $\mu_1$ computed by integrating equation~(\ref{eq:bower}).  Note
that for arbitrary initial conditions the EPS formalism cannot
analytically evaluate the higher moments of the distribution.
  
The agreement between PINOCCHIO and the simulations varies from the 5
per cent of the $\Lambda$CDM to the 10 per cent of the SCDM case but
it does not depend on the redshift.  Again PINOCCHIO is found to
improve with respect to EPS. In particular, at low redshift the EPS
predictions underestimate the mean value by a factor that ranges from
20 per cent to 30 per cent.

Our results can be compared to those shown by Sheth \& Lemson (1999b)
and Somerville et al. (2000) for EPS based merger trees and with Sheth
\& Tormen (2001) who elaborate an excursion set model based on
ellipsoidal collapse.  In general, PINOCCHIO reproduces the
statistical properties of progenitor distributions with better
accuracy then the other methods.  It is remarkable that the tests
based on parent haloes with different mass ranges give very similar
results, reproducing the simulations with a comparable accuracy.
\begin{figure}
\centerline{
\psfig{file=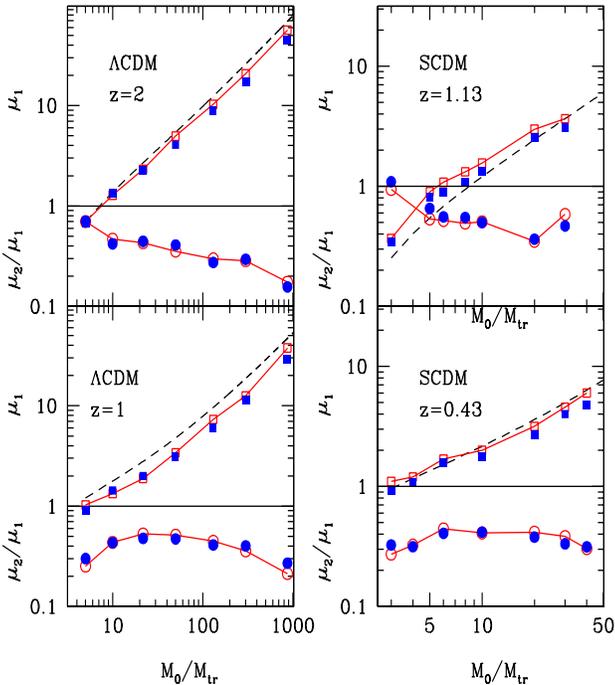,width=8.5cm,height=10cm}}
\caption{ The first two moments of the distribution of the number of
  progenitors $P(N,M_0)$ as a function of the parent mass $M_0$.  The
  left plots show the $\Lambda$CDM case at \rs $z=1$ and $2$.  The
  threshold mass is $M_{\rm th}=2.3 \times 10^{11} M_\odot $ (30
  particles) and we plot the mean (squares) and the rescaled variance
  (circles) up to $M_0=1000 M_{\rm th}$. The solid lines with open
  symbols are the PINOCCHIO results and the filled symbols are the
  simulation data.  The dashed line is the EPS analytical prediction
  for the mean.  The right plots show the SCDM case at \rs z=0.43 and
  z=1.13.  The threshold mass is $M_{\rm th}=1.3 \times 10^{14} M_\odot $
  (30 particles), and we plot the mean and the rescaled variance up to
  $M_0=50 M_{\rm th}$.}
\label{fig:meanL}
\end{figure}

\subsection{Object-by-object comparison}
We finally test the degree of agreement between PINOCCHIO and the
simulations at the object-by-object level for the number of
progenitors that are cleanly reconstructed.
In paper I and paper II a pair of haloes coming from the two catalogues
(PINOCCHIO and FOF) were defined as cleanly assigned to each other if
they overlapped in the Lagrangian space for at least 30 per cent of
their volume and no other object overlapped with either of them to a
higher degree.  The cleanly assigned haloes were shown to overlap on
average at the 60-70 per cent  over at all
redshifts.  The fraction of cleanly assigned haloes was found to
depend on the degree of non linearity reached by the system, decreasing
from almost 100 per cent to 70 per cent, at worst, at later times.
This is due to the lower accuracy of the Zel'dovich approximation in
predicting the displacements as the density field becomes more and
more non-linear (see paper II).

We now quantify the number of PINOCCHIO progenitors that are cleanly
assigned to FOF progenitors for each cleanly assigned parent halo.
For this analysis we restict ourselves to the $\Lambda$CDM case, which
gives a wider mass range but a higher level of non-linearity.
\begin{figure}
\centering
\psfig{file=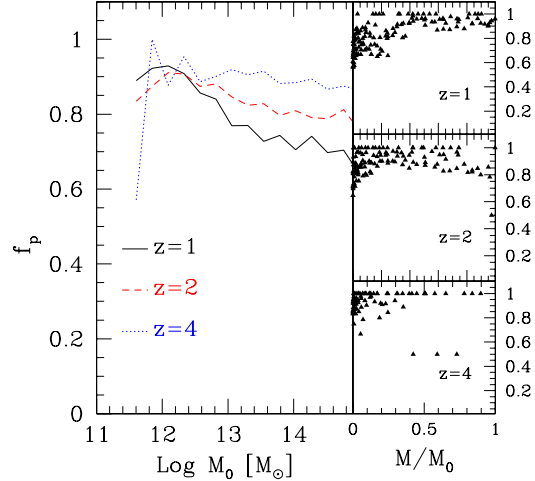,width=7cm,height=7cm}
\caption{ Fraction $f_{\rm p}$ of cleanly assigned progenitors  
  for \rs $z=1$, $2$ and $4$.  The left hand side represents the
  fraction of cleanly assigned progenitors as function of the parent
  mass. The three lines correspond to the different redshifts. The
  tree plots on the right are the scatter plots of $f_{\rm p}$ as function
  of the progenitor mass $M$ normalized to the parent mass $M_0$, for
  the parent haloes of mass $ 10^{11}$ M$_\odot$ $< M_0 < 10^{15}$
  M$_\odot$; the redshift increases from top to bottom.}
\label{fig:obyo}
\end{figure}

In Fig.~\ref{fig:obyo} we show, for the parents that are cleanly
identified, the fraction in number $f_{\rm p}$ of the progenitors that are
cleanly identified as well.  This quantity is shown both as a function
of the parent mass $M_0$ and as a function of the progenitor mass
$M/M_0$ in units of the parent mass.  The number of cleanly identified
progenitors ranges from 60 to 100 per cent, with an average value
between 80 and 90 per cent. The fraction $f_{\rm p}$ is in general
higher at higher redshift, when the object-by-object agreement between
PINOCCHIO and the simulation is better. As a function of $M_0$ larger
parent haloes tend to be reconstructed with worse accuracy, especially
at $z=1$.  This is mainly due to the small progenitors, as the right
panels of Fig.~\ref{fig:obyo} show. The progenitors which carry a mass
of less that $\sim$20 per cent are those that are worst reconstructed.
We conclude that PINOCCHIO is able to reconstruct correctly the main
branches of the merger trees, while secondary branches, especially
present in the larger haloes, are reconstructed in a noisier way.
\begin{figure*}
\centerline{ 
\psfig{file=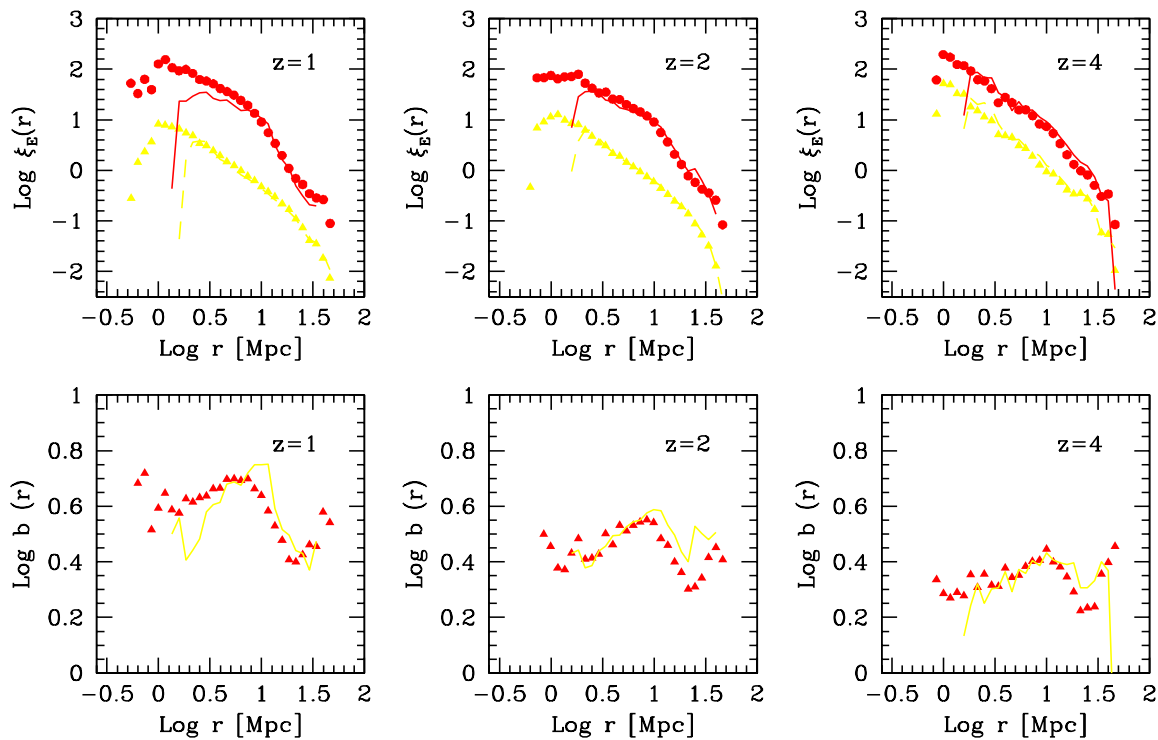,width=18cm,height=11cm}}
\caption{The Correlation functions in the $\Lambda$CDM case 
  for progenitors of mass greater than 10$^{14} M_\odot$ at redshift 0
  (circles and solid lines), compared with the the correlation
  function for all the haloes with mass greater than the threshold mass
  $M_{\rm tr}=2.3\,10^{11} M_\odot$ (triangles and dashed lines).
  Points refers to FOF selected haloes from the simulation and lines to
  PINOCCHIO haloes.  The second row of plots shows the ratio between the
  progenitors and total correlation functions.  The \rs increases from
  left to right and covers the values $z=1.0, \; 2.0,\; 4.0$.  }
\label{fig:xiL}
\end{figure*}

\section{The spatial properties of merging haloes}

One remarkable limit of EPS is the lack of spatial information for the
haloes.  Several authors (Mo \& White 1996; Mo, Jing \& White 1997;
Catelan et al. 1998; Porciani et al. 1998) found approximate
analytical expressions for the bias of haloes of fixed mass, i.e. for
the ratio between the two-point correlation function of haloes and that
of the underlying matter field.  Such analytical estimates have been
found to agree with the results of simulations to within $\sim$40 per
cent (Mo \& White 1996; Jing 1998; Porciani, Catelan \& Lacey 1999;
Sheth, Mo \& Tormen 2000; Colberg et al. 2001).  In this approach it is
not possible to know how the bias changes for haloes with different
merger histories.  This piece of information is precious to produce
predictions on the bias of galaxies of different types, that typically
have different merger histories.

As shown in paper I and II, PINOCCHIO haloes have the same correlation
lenght $r_0$ as FOF haloes to within 10 per cent error.  Having
knowledge of both merger histories and halo positions, PINOCCHIO can
provide information on the relation between clustering and merging.
To show this, we select PINOCCHIO and FOF haloes in the $\Lambda$CDM
cosmology at $z=0$ with masses greater than $10^{14}$ M$_\odot$. We
check their merging histories at $z=1$, $2$ and $4$, and we evaluate
the the two-point correlation functions for their progenitors.  In
Figure \ref{fig:xiL} the solid lines represent the two-point
correlation function of progenitors, $\xi_{\rm p}(r)$, evaluated in
PINOCCHIO compared with the same quantity measured in the simulation.
The plots show that PINOCCHIO reproduces such correlation functions to
within $\sim$20 per cent error.

We also compare this function with the average correlation function,
$\xi_{\rm h}(r)$, at the same redshifts.  It is apparent that
PINOCCHIO reproduces correctly the larger clustering amplitude of
haloes that flow into cluster sized one.  The bias between the two halo
populations is defined as: $b^2(r,z)=\xi_{\rm p}(r)/\xi_{\rm h}(r)$.
We compare in the bottom row of plots in Fig.~\ref{fig:xiL} the bias
measured in the simulation with PINOCCHIO results.  The bias is
recovered to within $\sim$20 per cent and the scale dependence is
correctly reproduced.

\section{Orbital parameters of the merging haloes}
In the hierarchical clustering scenario a merging event between two or
more haloes corresponds to the loss of identity of the single primitive
units which merge to form a new halo. However, high-resolution N-body
simulations show that the dynamical evolution after an encounter is
more complicated than this idealized picture: the haloes may retain
their identity, and become substructure of the new system (Moore,
Katz \& Lake 1996; Tormen 1997; Ghigna et al. 1998; Tormen, Diaferio
\& Syer 1998).  Indeed, this is in line with the same evidence of
galaxies within galaxy groups or clusters.  The life of these
substructures is affected by the varius dynamical effects that
contribute to their disruption. The dynamical friction force drives
the satellites towards the center of mass of the system where they can
merge with the central object or among themselves.  While a satellite
orbits inside the main halo, the tidal forces exerted by the
background induce its evaporation and reduce its mass (Gnedin \&
Ostriker 1997; Gnedin, Hernquist \& Ostriker 1999; Taylor \& Babul
2000; Taffoni et al. 2001, Taffoni et. al 2001 in prep).
 
The evolution of substructure is one of the crucial points to modeling
galaxy formation. A most important aspect of this is the prediction of
the initial orbital parameters, i.e. the energy and the angular
momentum of the orbit of the satellites infalling into the main halo.
The large-scale simulations as those we use in the present paper lack
enough resolution to address such processes.  High-resolution
simulation are necessary to describe the evolution of satellites
(Tormen 1997; Ghigna et al. 1998), at the cost of simulating one
cluster at a time.  It is then useful to resort again to analytic
modeling of the dynamical friction and tidal stripping (Chandrasekhar
1943; see e.g., Binney \& Tremain 1987, Lacey \& Cole 1993; van den
Bosch et al. 1999, Colpi, Mayer \& Governato 2000).  These analytical
models require knowledge of the initial orbital parameters. In the
semi-analytical, EPS-based codes for galaxy formation, these
parameters are in general Monte-Carlo extracted from some
distributions obtained from high-resolution simulations (Tormen 1997;
Ghigna et al.  1998).
%
\begin{figure}
\centerline{
\psfig{file=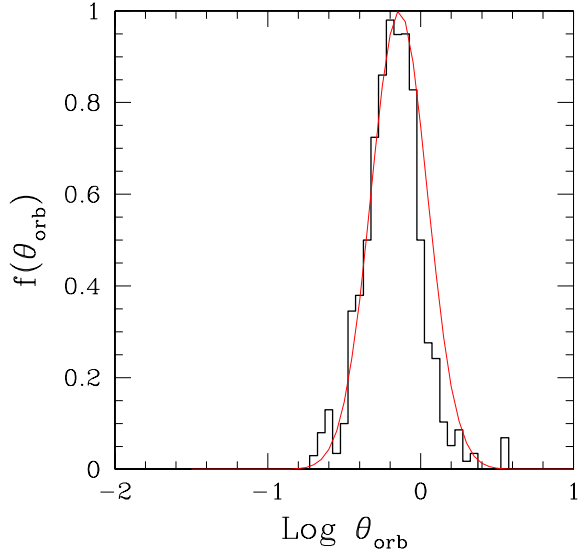,width=10cm,height=10cm}}
\caption{The distribution of the $\Theta_{\rm orb}$ for the satellites
  that merge with a halo of mass $M=2\times 10^{14}$~M$_\odot$ at
  z=0.}
\label{fig:ecc}
\end{figure}

Within the PINOCCHIO code, it is possible to predict the impact
parameters of the merging satellites, as the infall velocities and the
relative distances are known. Notice that this calculation is
analogous to that of angular momentum of haloes presented in paper II.
Given the impact (Zel'dovich) velocity $\Delta \bld v$ and the
relative distance $\Delta r$ the angular momentum and the energy are
computed as: \3
{\bf J} &=& \Delta {\bf r} \wedge \Delta {\bf v} \\
E&=&\frac 1 2 \left(\Delta {\bf v} \right)^2+\phi(|\Delta {\bf r}|)\;.
\4 $\phi(|\Delta \bf {r}|)$ can be evaluated as the gravitational
potential of a point mass which touches the external layer of a
spherical halo of mass $M$: $\phi(|{\bf r}|)=GM/|{\bf r}|$. The linear
growth of the relative velocity is stopped at a physical time equal to
a half of the merging time (see paper II).

To study the ability of PINOCCHIO in predicting the orbital parameters
of DM substructures we compute the distribution of the orbital
parameters of the satellites that merge with a halo of mass $M=2
\times 10^{14}$~M$_\odot$.  We express the angular momentum and the
energy for unit mass in terms of the circularity $\epsilon \equiv
J/J_{\rm c}(E)$, where $J_{\rm c}(E)=V_c \; r_{\rm c}(E) $ is the
angular momentum and $r_{\rm c}(E)$ is the radius of the circular
orbit with the same energy (see eg., van den Bosch et al. 1999).  To do
that we assume a Navarro, Frank \& White (1996) density profile for
the main halo and we calculate the associted potential energy profile
$\phi_{\rm NFW}(R)$ and the associated circular velocity profile
$V_{\rm c}(R)$ (see e.g. Navarro, Frank \& White 1996; Klypin et al.
1999, Taffoni et al. in prep).  We consider a particular combination
of the orbital parameters
\1
\Theta_{\rm orb}=\epsilon^{0.78}\,\left[{r_{\rm c}(E) \over R_{\rm vir}}
\right ] ^2 \;,
\2
introduced by Cole et. al (2001). They use the Tormen (1997) results
to derive the distribution for the $\Theta_{\rm orb}$ factor and they
find that this distribution can be fitted with a log normal function
with mean value $\langle\log_{10}(\Theta_{\rm orb})\rangle=-0.14$ and
dispersion $\langle(\log_{10}(\Theta_{\rm
  orb})-\langle\log_{10}(\Theta_{\rm
  orb})\rangle)^2\rangle^{0.5}=0.26$.
 
The result of our analysis are presented in Fig.~\ref{fig:ecc}, we
compare the distribution of the $\Theta_{\rm orb}$ factor measured in
PINOCCHIO (histogram) with the theoretical fit derived by Cole et. al
(2001) (solid line).  We note that the distribution measured form
PINOCCHIO reproduce with good accuracy the log normal function.  The
average value derived by our analysys is $\langle\log_{10}(\Theta_{\rm
  orb})\rangle=-0.18$ and the dispersion is
$\langle(\log_{10}(\Theta_{\rm orb})-\langle\log_{10}(\Theta_{\rm
  orb})\rangle)^2\rangle^{0.5}=0.23$.

\section{Conclusions}
We have tested the predictions of the PINOCCHIO code, presented in
paper I and paper II, regarding the hierarchical nature of halo
formation, with particular attention to those aspects that are mostly
relevant for galaxy formation. We have compared the results of
PINOCCHIO with those of two large N-body simulations ($\Lambda$CDM and
SCDM cosmologies)  drawing the following conclusions:
\par\noindent 
1. The merger histories of the PINOCCHIO haloes resemble closely those
found applying the FOF algorithm to the N-body simulations.  The
agreement is valid at the statistical level for groups of at least 30
particles (good results are obtained even for haloes of 10 particles).
\par\noindent 
2. Statistical quantities like the conditional mass function, the
distribution of the largest progenitor, the ratio of the second
largest to largest progenitors, and the higher moments of the
progenitor distributions are recovered with a typical accuracy of
$\sim$20  per cent.  
\par\noindent 
3. The agreement is good also at the object-by-object level, as
PINOCCHIO cleanly reproduces $\ga$70 per cent of the progenitors when
parent haloes are cleanly recognised themselves.  The agreement slowly
degrades with time. 
\par\noindent 
4. The increased noise recovered in the object-by-object
agreement, as time progresses and non-linearity grows, does not
influence the accuracy of the predictions in a statistical sense.
\par\noindent 
5.  The correlation function of higher-redshift haloes that are
progenitors of lower-redshift massive haloes is correctly reproduced
to within an accuracy of $\sim$10 per cent in $r_0$. The scale
dependent bias of these with respect to the total halo population is
also reproduced to within an accuracy of 20 per cent or better.
\par\noindent 
6. PINOCCHIO gives an estimate of the initial orbital parameters
of merging haloes as well, which is found in reasonable agreement with
available results from high-resolution N-body simulations.

Compared with the widely used EPS approach (Bond et al. 1991; Bower
1991; Lacey \& Cole 1993), PINOCCHIO improves in most respects.
\par\noindent
1. The fit of the statistical quantities achieved by PINOCCHIO is much
better than the PS and EPS estimates, which show discrepancies up to a
factor of 2 (see Governato et al. 2000; Jenkins et al. 2001; Bode et
al. 2001; Somerville et al. 1999; Sheth \& Lemson 1999b, Bagla et al.
199?).
\par\noindent
2. The validity of PINOCCHIO extends to the object-by-object level, in
contrast to EPS (Bond et al. 1991; White 1996).
\par\noindent
3. PINOCCHIO is not affected by the inconsistencies of the EPS
approach, that can be corrected only by means of heuristical recipes
(Somerville \& Kolatt 1999; Sheth \& Lemson 1999; Cole et al. 2001).
\par\noindent
4. PINOCCHIO provides much more useful information on the haloes, such
as positions, velocities and angular momenta ad initial orbital
parameters at merger; at the same time it is not more computational
demanding than an EPS based code to generate the merging histories of
haloes.

These results confirm the validity of PINOCCHIO as a fast and flexible
tool to study galaxy formation or to generate catalogues of galaxies
or galaxy clusters, suitable for large-scale structure studies. In
fact PINOCCHIO reproduces, in a much quicker way and to a very good
level of accuracy, all the information that can be obtained from a
large-scale N-body simulation with pure dark matter, without needing
all the post processing necessary to obtain the merger trees of haloes.

PINOCCHIO is available at http://www.daut.univ.trieste.it/\~\ pinocchio.

\section*{Acknowledgments}

We thank Stefano Borgani, Monica Colpi, Fabio Governato, Lucio Mayer
and Cristiano Porciani for useful discussions. GT thanks Valentina
D'Odorico for her critical reading of the manuscript.  TT thanks PPARC
for the award of a post-doctoral fellowship.  N-body simulations were
run at the ARSC and Pittsburg supercomputing centres.  Research
conducted in cooperation with Silicon Graphics/Cray Research utilising
the Origin 2000 super computer at DAMTP, Cambridge.

\label{lastpage}
\end{document}